\documentstyle[aps,prl,twocolumn]{revtex}

\begin{document}

\draft

\title{Convergent expansions for properties of the Heisenberg model
for CaV$_{\bbox{4}}$O$_{\bbox{9}}$}

\author{M. P. Gelfand,$^1$\cite{byline1}  Zheng Weihong,$^2$\cite{byline2}
Rajiv R. P. Singh,$^3$\cite{byline3} J. Oitmaa,$^2$\cite{byline4}
and  C. J. Hamer$^2$\cite{byline5} }
\address{${}^1$Department of Physics,
Colorado State University,
Fort Collins, Colorado 80523\\
${}^2$School of Physics,                                              
The University of New South Wales,                                   
Sydney, NSW 2052, Australia\\                      
${}^3$Department of Physics,                                              
University of California,                                   
Davis, California 95616.}

\maketitle 

\begin{abstract}
We have carried out a wide range of calculations for the $S=1/2$ 
Heisenberg model 
with nearest- and second-neighbor interactions
on a two-dimensional lattice which describes the geometry of the vanadium
ions in the spin-gap system CaV$_4$O$_9$.
The methods used were  convergent high-order perturbation
expansions (``Ising'' and ``Plaquette'' expansions at $T=0$,
as well as high-temperature expansions) for quantities
such as the uniform susceptibility, sublattice magnetization,
and triplet elementary excitation spectrum.
Comparison with the data for
CaV$_4$O$_9$ indicates that its magnetic properties are well described by
nearest-neighbor exchange of about 200K in conjunction with
second-neighbor exchange of about 100K.

\end{abstract}                                                              

\narrowtext
Since the discovery of high temperature superconductivity in the
cuprates there has been much interest in two-dimensional antiferromagnetism.
From a theoretical point of view, there has been an extensive search
for the so-called spin liquid ground state
in models with realistic interactions. 
A possible link between such a ground state and high temperature
superconductivity has often been suggested though not demonstrated.
For square lattice Heisenberg models
it is believed that sufficient further-neighbor interactions
will destabilize the N\'eel state and
lead to a spin-gap phase; however, such a phase typically has
spin-Peierls order and hence is not a true spin liquid \cite{dimer}. 
A spin-gap phase without other spontaneously
broken symmetries has been found experimentally in 
quasi-one-dimensional spin ladders \cite{ladder} and 
theoretically in strongly coupled two-plane models \cite{2plane}. 

In light of these developments, it is
interesting that the quasi-two-dimensional Heisenberg system
CaV$_4$O$_9$ has been found
to exhibit a spin gap \cite{taniguchi}.
Within each layer, the spins form a one-fifth depleted square
lattice, which we denote the CAVO lattice (see Fig.~\ref{CAVOlattice}). 
Recently this system has attracted much attention from theorists
\cite{ueda,katoh,sano,mila,troyer,starykh}.
It is now believed that the spin degrees of freedom in CaV$_4$O$_9$ 
are in the ``plaquette phase,'' that is, there would be no phase
transition if the intact squares (plaquettes) of the CAVO lattice
were adiabatically decoupled.
It has been argued \cite{ueda} that the superexchange between these spins is
mediated by oxygen atoms, which are situated out of the plane at
the center of the squares in the CAVO lattice, 
and that this may lead to significant
second neighbor interactions: 
as a first approximation, one expects $J_1=J'_1=2J_2=2J'_2$. 
Furthermore, it has been argued that due to its special structure 
this lattice is susceptible to symmetry-preserving lattice 
distortions \cite{starykh}. 
Hence it is appropriate to study a Heisenberg model with interactions
\begin{eqnarray}
{\cal H}= && J_1\sum_{(i,j)} {\bf S}_i \cdot  {\bf S}_j+
J'_1\sum_{(i,k)} {\bf S}_i \cdot  {\bf S}_k \nonumber\\
&&+J_2\sum_{(i,l)} {\bf S}_i  \cdot {\bf S}_l +
J'_2\sum_{(i,m)} {\bf S}_i  \cdot {\bf S}_m,
\label{Ham}
\end{eqnarray}
where the first sum runs over nearest-neighbor bonds within plaquettes,
the second over nearest-neighbor bonds between plaquettes, 
the third over second-neighbor bonds within plaquettes,
and the last over second-neighbor bonds between plaquettes.

We have carried out a variety of high-order convergent perturbation
expansions for this Hamiltonian that are comparable in
accuracy to the best available quantum Monte Carlo calculations (which
have been limited to $J_2=J'_2=0$) and that provide the only reliable
results for the parameter values most relevant to CaV$_4$O$_9$.
An outline of the calculations and results is as follows:
(1) Ising expansion estimates of the sublattice magnetization and 
phase boundary for models
with $J_2=J'_2=0$ are consistent with the quantum Monte Carlo
calculations of Troyer {\it et al.} \cite{troyer};
(2) Plaquette expansions for the triplet elementary excitation 
spectrum are highly convergent and informative in the plaquette
phase, and (3) High-temperature 
expansions for the susceptibility, combined with plaquette expansion
results, can be compared with the existing
experimental data \cite{taniguchi} to fix the model's parameters
for CaV$_4$O$_9$.  In this Letter we can only describe the highlights
of the results; details are reserved for a separate publication.

For the Ising expansions we have considered the Hamiltonian
\begin{eqnarray}
\label{Ising}
{\cal H}&&=J_1\sum_{(i,j)} [S_i^z S_j^z+\alpha (S_i^x S_j^x+S_i^yS_j^y)] \\
+&&J'_1\sum_{(i,k)} [S_i^z S_k^z +\alpha (S_i^x S_k^x+S_i^yS_k^y)]
+t(1-\alpha)\sum_i \epsilon_i S_i^z, \nonumber
\end{eqnarray}
where $\epsilon_i=\pm 1$ on the two sublattices. 
(The last term is a local field term, which can be included
to improve convergence and proves useful in that regard when $J'_1\approx J_1$.)
We have developed expansions up to order
$\alpha^{14}$, for several values of $J'_1/J_1$,
for the ground state energy, the sublattice magnetization $M$, and
the longitudinal and transverse susceptibilities. The expansion
techniques have been reviewed earlier \cite{he90,gel90}.  
When the Heisenberg model
has a spin gap, one expects these expansions to break 
down at some $\alpha<1$ and
the system to exhibit $d=3$ Ising criticality.
Such considerations can be used to estimate the domain of 
long-range N\'eel order.
Within the N\'eel-ordered phase our estimates of $M$ 
are presented in Fig.~\ref{Mfig} along with those
of Troyer {\it et al.} \cite{troyer}.

For the plaquette expansions we return to the original
Hamiltonian of interest, Eq.~(\ref{Ham}), but with the
interplaquette interactions parametrized as $J'_1=\lambda_1 J_1$,
$J_2=\gamma J_1$,
and $J'_2=\lambda_2 J_2$.  For convenience we have taken
$\lambda_1=\lambda_2\equiv\lambda$ in our calculations;
that restriction could be relaxed if desired.
Let us restrict our attention to $\gamma < 1$, since there is both
a ground state and lowest excited state level crossing for an
isolated plaquette when $J_1=J_2$, and $J_1>J_2$ is certainly
more relevant to CaV$_4$O$_9$.
When $\lambda=0$ the ground state is a product of 
singlets on the non-interacting plaquettes, while
the lowest-lying triplet excitations are gapped and completely local.
For $\lambda\ne 0$, the excitations are mobile and develop a
well defined dispersion.
Using a recently developed method for computing the 
excitation spectra \cite{gelfand},
the complete dispersion for these triplet excitations 
has been constructed to
order $\lambda^5$ for various values of $\gamma$. 
Following the convention of Katoh and Imada \cite{katoh} for the wave vectors,
to second order in $\lambda$ the dispersion has the form
\begin{eqnarray}
\Delta(q_x,q_y)/J_1 = 1+\lambda^2\,c_2^{(0,0)}(\gamma) \nonumber\\
\qquad + \textstyle{1\over2}\lambda(\cos q_x + \cos q_y)[c_1^{(1,0)}(\gamma)
                               +\lambda\,c_2^{(1,0)}(\gamma)]
\nonumber\\
\qquad + \lambda^2\cos q_x \cos q_y\,c_2^{(1,1)}(\gamma) \nonumber\\
\qquad + \textstyle{1\over2}\lambda^2(\cos2q_x + \cos2q_y)\,c_2^{(2,0)}(\gamma)
\label{secondorderdisp}
\end{eqnarray}
The coefficients $c_n^{\bf r}$ for $\gamma=0$, 0.4 and 0.5 are presented in Table~\ref{smallpexp}.
(Because $\gamma$ is a parameter in both the unperturbed and perturbing
Hamiltonians, a simple double expansion in which $c_n^{\bf r}$ can be expressed
as a polynomial of order $n$ in $\gamma$ does not exist.)
Second-order excitation-spectrum expansions have been presented earlier:
for $J_2=0$ by Katoh and Imada \cite{katoh} 
and at $q_x=q_y=\pi$  for $J_2=J'_2$  (that is, $\lambda_1$ and $\gamma$ arbitrary
and $\lambda_2=1$)
by Ueda {\it et al.} \cite{ueda}.
However, in both cases the authors failed to consider the 
$O(\lambda^2)$ term associated with the $(\cos q_x + \cos q_y)$
part of the dispersion; in effect they set $c_2^{(1,0)}$ to zero.
For the case $\lambda=1$ and $\gamma=0$
one finds from summing the correct series terms to $O(\lambda^2)$ a gap 
$\Delta(\pi,\pi)=0.010 J_1$,
in contrast to the previous perturbative estimates of $0.205 J_1$. As evidenced
by the Ising expansions and the Monte Carlo results of Troyer
{\it et al.}, this point lies in the N\'eel-ordered phase and should have a
zero gap. This shows that already in second order the plaquette expansion for the gap
locates the phase boundary fairly well, at least for $\gamma\approx0$.

Another important point not discussed in earlier 
perturbation-theoretic work is that
the minimum of $\Delta$ is not necessarily found at ${\bf q}=(\pi,\pi)$.
For small $\lambda$ the minimum shifts from $(\pi,\pi)$ when 
$\gamma<1/2$ to $(0,0)$ when $\gamma>1/2$.  Precisely at $\gamma=1/2$
the minimum for small $\lambda$ is achieved at $(\tilde q,\tilde q)$ with
$\tilde q/\pi = 0.7044\ldots$.
  To the extent one can trust estimates
of $\Delta$ based on simply summing series coefficients, 
such an incommensurate  minimum persists at finite $\lambda$ and for
$\gamma\approx 1/2$.  (A similar incommensuration for sufficiently
large $\gamma$ was found in the mean-field calculations of Starykh
{\it et al.}~\cite{starykh}.)

The present plaquette expansions are also the first which are sufficiently long to reflect
the absence of reflection symmetry in the Hamiltonian: in third order
and above $\Delta(q_x,q_y)\neq\Delta(q_y,q_x)$.  
It is not a dominant feature of the excitation spectrum but 
if $\gamma$ is sufficiently large it may be observable:
at $\gamma=0.5$ and $\lambda=1$ the asymmetry parameter 
\begin{equation}
2[\Delta(q,\pi-q)-\Delta(\pi-q,q)]/[\Delta(q,\pi-q)+\Delta(\pi-q,q)]
\nonumber
\end{equation}
may be 10\% for $q/\pi\approx 0.15$.  

The fact that the excitation spectrum  minimum is not fixed at 
${\bf q}=(\pi,\pi)$
could in principle complicate the matter of estimating the gap
in the plaquette phase.  
However, for $\gamma\leq 1/2$ and $\lambda \leq 1$ the value
of $\Delta(\pi,\pi)$ exceeds the true gap by at most $0.05 J_1$
and usually by far less.  So it is not too bad to identify estimates of
$\Delta(\pi,\pi)$ with estimates of the gap 
in the parameter region of most physical interest.
Another consideration is that the uncertainty
in the estimates of $\Delta(\pi,\pi)$, which can be judged by
spread in the direct sums and Pad\'e approximants, become noticeable precisely
in the part of $\gamma$-$\lambda$ space where the gap minimum
apparently moves away from ${\bf q}=(\pi,\pi)$.
See Fig.~\ref{Deltaplot} for some examples of $\Delta(\pi,\pi)$ estimates.
Fortunately, for the parameters expected to be relevant to CaV$_4$O$_9$
the estimates of the gap appear to be fairly reliable, if not
quite as good as when $\gamma \approx 0$.
For those parameters,  $J'_1=J_1=2J'_2=2J_2$ ($\lambda=1$, $\gamma=1/2$),
the gap is $0.57\pm 0.03 J_1$.
Taking the experimental gap value of $107\,$K
for CaV$_4$O$_9$ 
leads to the estimate $J_1=190\pm10\,$K.
The Curie temperature ($=9J_1/8$, for the given
parameter ratios) should then be $215\pm10\,$K, which is consistent
with the experimental value of $220\,$K and shows that the present parameter
estimates are entirely plausible.

If magnetic neutron scattering data can be obtained, comparison
with the plaquette expansion for the gap should allow for precise
determination of model parameters.  The structure of the triplet
excitation spectrum is strongly dependent on the ratio of nearest-
to second-neighbor couplings:
see Fig.~\ref{dispplot} for some illustrative plots.

We now turn to the high temperature expansions for the uniform
susceptibility.
The uniform susceptibility per spin can be defined as
$\beta \chi(T)= {1\over N}\sum_i\sum_j \langle S^z_i S^z_j \rangle$
where the angular brackets represent thermal averaging and $\beta=1/T$.
For $J_2=J'_2=0$ we have determined the expansion to order $\beta^{14}$. 
The integrated differential approximants
have been used to extrapolate $\chi$
to low $T$; we find that they agree with Monte Carlo
calculations \cite{troyer} down to  $T\approx J_1/10$ when the
ground state lies in the plaquette phase.

High temperature expansions have also been developed for the models 
with $J_1=J'_1$, $J_2=J'_2\ne 0$,
to order $\beta^{10}$. They should be more relevant for comparison with the
CaV$_4$O$_9$ data, however, they have proven more difficult to extrapolate
to lower temperatures. In order to improve the convergence, we have
developed differential approximants for $\chi e^{\Delta/T}$, with $\Delta$
taken from our studies at $T=0$. This biases the leading exponential
behavior of the susceptibility at low temperatures to be of the form
$e^{-\Delta/T}$. Despite this biasing there remain significant
uncertainties in the location of the peak temperature, where $\chi$ is
maximized;
the best we can say is that 
it lies in the range $0.4$--$0.6\ J_1$ for $J_2=J_1/2$.
The susceptibility of CaV$_4$O$_9$
peaks at about $110\,$K, which is consistent with a $J_1$ 
value of approximately $200\,$K.

A comparison of experimental data with the calculated susceptibility, 
using $J_1=190\,$K as estimated above,
is shown in Fig.~\ref{chifig}. In order to convert the susceptibility
for the Heisenberg model into units of emu/g, we multiply the theoretical
value
(calculated with $J_1=1$) by
$(g^2\mu_B^2/J_1 k_B) (4N_A/M)$, where $\mu_B$ is the Bohr
magneton, $k_B$ is the Boltzmann constant, $N_A$ is Avogadro's number,
$M$ is gram molecular weight, and the factor
of 4 comes from the four vanadium spins per CaV$_4$O$_9$ formula unit. 
The $g$-factor can be used as a fitting parameter, and the comparison shown 
uses the value 1.77. We note that the calculated susceptibility does
not fully agree with the data at high temperatures: this may be in
part due to the existence of other phases above $340K$ as noted
in the experimental paper \cite{taniguchi}, 
or due to the lattice distortions discussed
by Starykh {\it et al.}~\cite{starykh}, which can make 
the exchange constants temperature
dependent. A more detailed examination of the high temperature susceptibility
for CaV$_4$O$_9$
in context of the Heisenberg models will be given elsewhere.

In summary, we have carried out a wide range of high-order perturbation
expansions for $S=1/2$ Heisenberg antiferromagnets on the CAVO lattice.
The existing susceptibility data for CaV$_4$O$_9$ is
generally consistent 
with the coupling ratios $J_1=J'_1=2J_2=2J'_2$ suggested by 
Ueda {\it et al.}~\cite{ueda}.
Measurements of the wave vector resolved triplet excitation spectrum 
should be compared with estimates of the spectrum based on the present
plaquette expansions for a definitive test of the model.

Acknowledgments.  This work has been supported by the National Science 
Foundation
under Grant Nos.~DMR 94--57928 (MPG) and DMR 93--18537 (RRPS).  
The work in UNSW is supported by a grant 
from the Australian Research Council. 
We would also like to thank Dr. Troyer for providing us with
the Monte Carlo data and to Dr. Sato for the experimental data
on CaV$_4$O$_9$.


\begin{figure}
\caption{The CAVO lattice, with sites indicated by circles.
The couplings $J_1$, $J'_1$, $J_2$, and $J'_2$ are indicated
by thick solid, thick dashed, thin solid, and thin dashed
lines, respectively.  Note that the plaquette centers lie
on a square lattice with spacing $b$ which is $\protect\sqrt 5$ times the
distance between nearest-neighbor sites.  In characterizing
the excitation spectrum (see Eq.~(\protect\ref{secondorderdisp}) below)
we take $b=1$ and rotate the coordinate system
so that the lines between nearest-neighbor plaquette centers
define the $x$ and $y$ axes.}
\label{CAVOlattice}
\end{figure}

\begin{figure}
\caption{Sublattice magnetization {\it versus\/} $J'_1/J_1$ (with
second-neighbor couplings $J_2=J'_2=0$) as estimated by 
Ising expansions (filled symbols) and 
the Troyer {\it et al.}\ \protect\cite{troyer}
quantum Monte Carlo calculations (open symbols).}
\label{Mfig}
\end{figure}

\begin{figure}
\caption{The excitation energy $\Delta(\pi,\pi)$ 
{\it versus\/} $\lambda$ for several values of $\gamma$.
}
\label{Deltaplot}
\end{figure}

\begin{figure}
\caption{The triplet excitation spectrum $\Delta(q_x,q_y)$ 
along high-symmetry cuts through
the Brillouin zone for coupling ratios $\lambda=1$ and 
$\gamma=0.1$, 0.3 and 0.5 (from greatest to least values of $\Delta(0,0)$).
The ends of the bars indicate the values of the direct sums to fourth 
and fifth order in $\lambda$.}
\label{dispplot}
\end{figure}

\begin{figure}
\caption{A comparison of the calculated susceptibility for the
Heisenberg model with $J_2=J_1/2$ (the solid lines representing
various approximants), with the experimental
data of Taniguchi {\it et al.}\ \protect\cite{taniguchi}
for CaV$_4$O$_9$.}
\label{chifig}
\end{figure}

\begin{table}
\caption{
Coefficients in the expansion of the triplet elementary excitation spectrum
$\Delta(q_x,q_y)$, to second order in $\lambda$, for selected $\gamma$ values:  
see Eq.~(\protect\ref{secondorderdisp}).
}
\begin{tabular}{crrr}
   $\ $   & $\gamma=0$\phantom{.000} & $\gamma=0.4$\phantom{00} & $\gamma=0.5$\phantom{00} \\
\hline
 $c_2^{(0,0)}$ & $-$0.0358796 & $-$0.1950309 &$-$0.2279762 \\
 $c_1^{(1,0)}$ &    0.6666666 &    0.1333333 &   0.\phantom{0000000} \\
 $c_2^{(1,0)}$ &    0.1944444 &    0.1799242 &   0.1597222 \\
 $c_2^{(1,1)}$ & $-$0.1111111 & $-$0.0444444 &   0.\phantom{0000000} \\
 $c_2^{(2,0)}$ &    0.0185185 &    0.0570370 &   0.0666667 \\
\end{tabular}
\label{smallpexp}
\end{table}

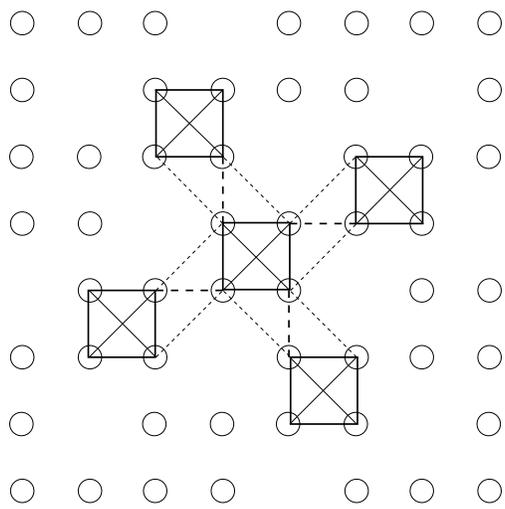
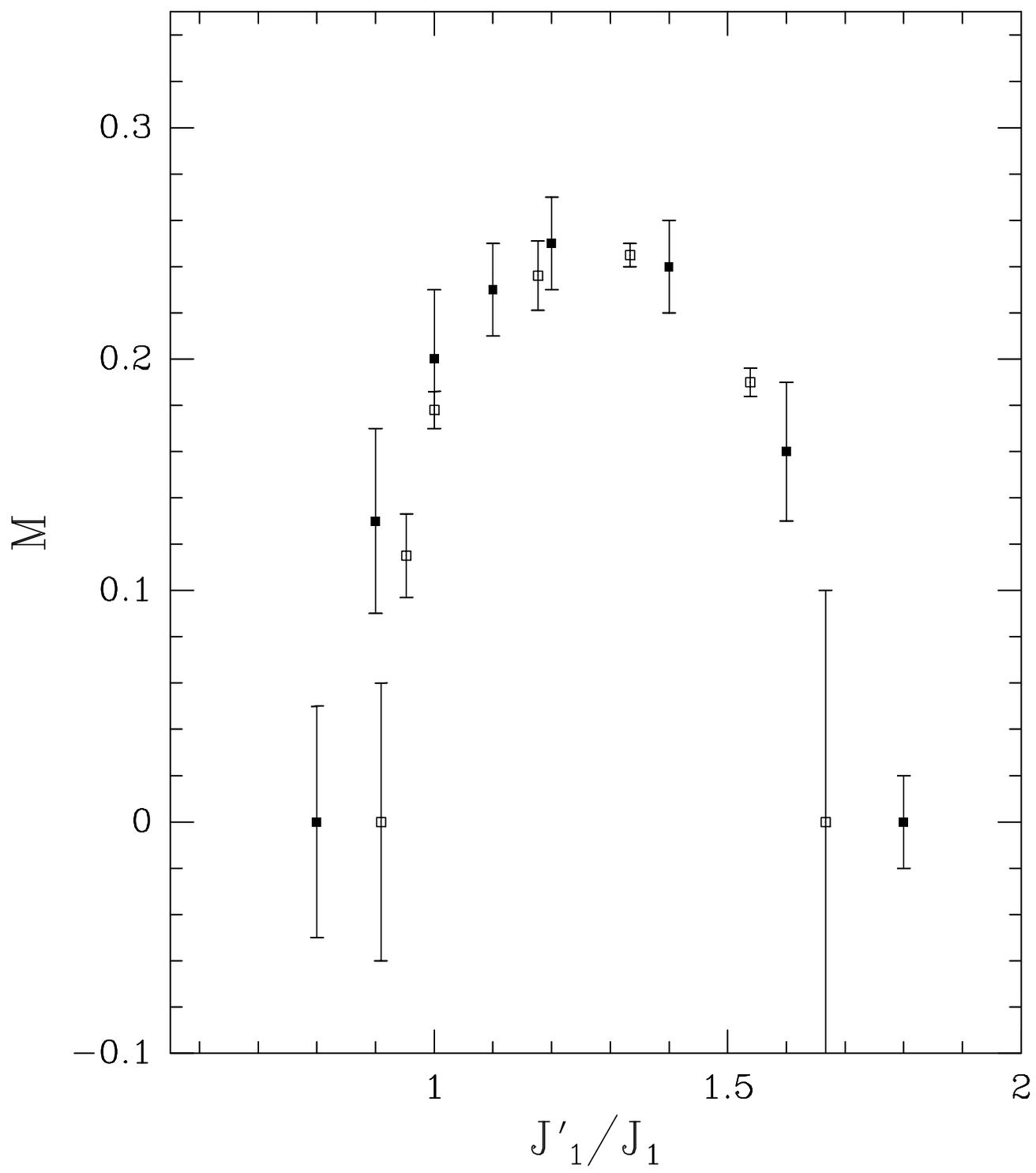
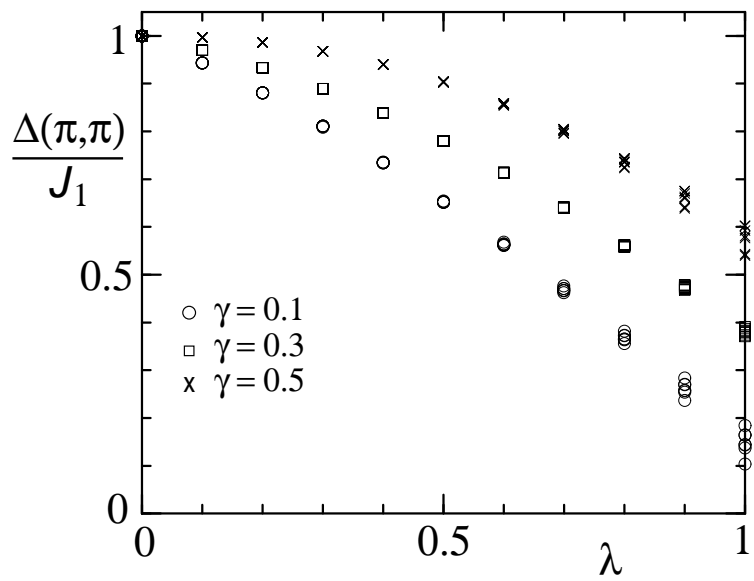
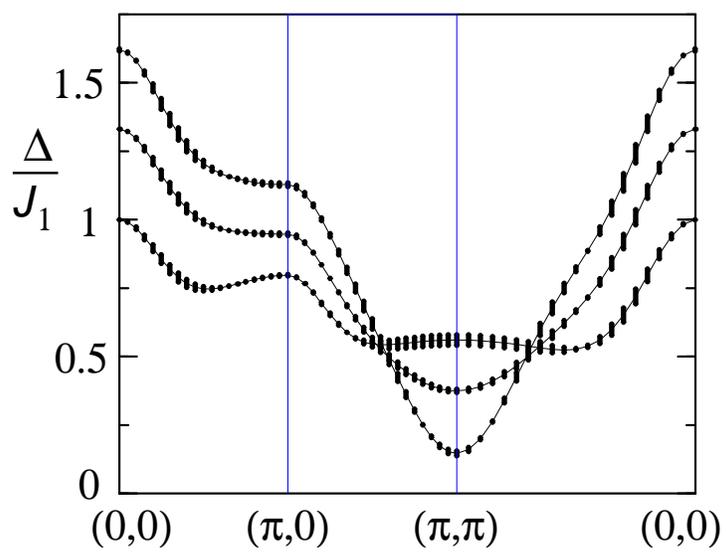
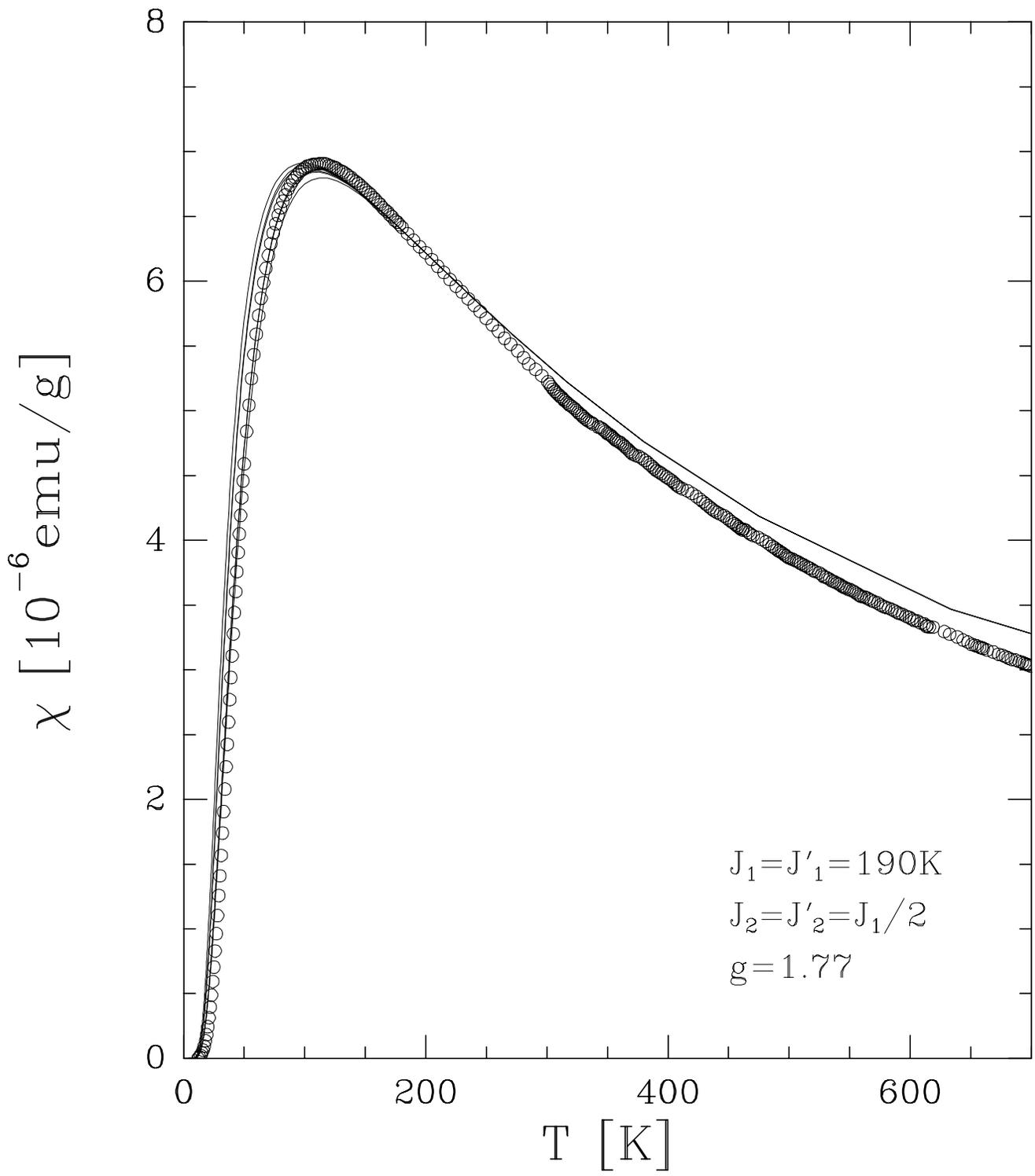
\end{document}